# NON-GAUSSIANITY OF MULTIPLE PHOTON SUBTRACTED THERMAL STATES IN TERMS OF COMPOUND-POISSON PHOTON NUMBER DISTRIBUTION PARAMETERS: THEORY AND EXPERIMENT

G. V. AVOSOPIANTS,[1,2,6] K. G. KATAMADZE,[1,2,3,4,6*] YU. I. BOGDANOV,[3,4,5,6] B. I. BANTYSH,[3,6] AND S. P. KULIK [1,2,4]

[1]*Quantum Technology Center of M. V. Lomonosov Moscow State University, 119991, Moscow, Russia*
[2]*Faculty of Physics of M. V. Lomonosov Moscow State University, 119991, Moscow, Russia*
[3]*Institute of Physics and Technology, Russian Academy of Sciences, 117218, Moscow, Russia*
[4]*A. M. Prokhorov General Physics Institute, Russian Academy of Sciences, Moscow, Russia*
[5]*National Research Nuclear University "MEPHI", 115409, Moscow, Russia*
[6]*National Research University of Electronic Technology MIET, 124498, Moscow, Russia*
*\*Corresponding author: k.g.katamadze@gmail.com*

**Annotation**

The multiphoton-subtracted thermal states are an interesting example of quantum states of light which are both classical and non-Gaussian. All the properties of such states can be described by just two parameters of compound-Poisson photon number distribution. The non-Gaussianity dependency on these parameters has been calculated numerically and analytically. The loss of non-Gaussianity during the optical damping has been also studied experimentally.

## 1. Introduction

Non-Gaussian quantum states – the states with non-Gaussian Wigner functions – as well as non-Gaussian operators, which transform Gaussian quantum states to the non-Gaussian ones, play a significant role in quantum optics. The non-Gaussianity is usually connected with non-classicality. Moreover, non-Gaussian quantum states or non-Gaussian quantum operations are necessary components for continuous-variable entanglement distillation [1–7]. Also non-Gaussian states enable significant improvement in quantum teleportation [8–10] and cloning [11] protocols.

Since non-Gaussianity as well as non-classicality serves as a resource for quantum information tasks, it is important to understand how to maintain it. It is well-known, that non-classicality of quantum states dramatically degrades under the action of optical losses. Recently, it was theoretically shown by the example of multi-photon subtracted thermal states (MPSTSs), that the non-Gaussianity exhibits the same behavior [12].

MPSTSs are both classical and non-Gaussian, so they are useful for studying non-Gaussianity separately from the non-classicality. Moreover, these states provide a good testing area for some fundamental theoretical problems. For example, photonic Maxwell's Demon have been demonstrated recently [13]. Also, as will be described below, non-Gaussianity of such states is related to entropy decrease, which can be considered as a source of work and information [14].

The paper is organized as follows. In Sec. 2 we give a conformity between the descriptions of MPSTSs, based on the Gauss hypergeometric function [13] and on the compound-Poisson distribution [15–17]. In Sec. 3 we show the dependences of several non-Gaussianity measures on the MPSTSs parameters. In Sec. 4 we describe the experimental technique for measurement of non-Gaussianity evolution during the optical damping. Finally, the experimental results are presented and discussed in Sec.5.

## 2. Multi-photon subtracted thermal states

The density matrix of a single-mode thermal state has a well-known diagonal form in a Fock basis:

$$\hat{\rho}_{TS}(\mu_0) = \sum_{n=0}^{\infty} P_{\mu_0}(n)|n\rangle\langle n|, \qquad (1)$$

where $P_{\mu_0}(n) = \mu_0^n / (1+\mu_0)^{n+1}$ – the Bose-Einstein distribution, and the parameter $\mu_0$ is the mean photon number. An *M*-photon subtracted state $\hat{\rho}_{MPSTS}$ is obtained by the multiple action of the annihilation operator $\hat{a}$ on $\hat{\rho}_{TS}$:

$$\hat{\rho}_{MPSTS} = \frac{\hat{a}^M \hat{\rho}_{TS}(\hat{a}^\dagger)^M}{\text{Tr}\left[\hat{a}^M \hat{\rho}_{TS}(\hat{a}^\dagger)^M\right]}, \qquad (2)$$

It turns out that both the thermal state and MPSTS can be described by the compound-Poisson photon-number distribution [15–17]:

$$\hat{\rho}_{MPSTS}(\mu,a) = \sum_{n=0}^{\infty} P_{\mu,a}(n)|n\rangle\langle n|,$$
$$P_{\mu,a}(n) = \frac{\Gamma(a+n)}{\Gamma(a)} \frac{\mu^n}{a^n n!} \frac{1}{\left(1+\frac{\mu}{a}\right)^{n+a}}. \quad (3)$$

Here $\Gamma(a)$ – gamma function, $a = M+1$ – coherence parameter, $\mu$ – mean photon number of MPSTS, which is connected with the mean photon number of initial thermal state $\mu_0$ by the equation $\mu = \mu_0(M+1)$. At $a = 1$ equation (3) turns to the Bose-Einstein distribution and at $a \to \infty$ – to the Poisson distribution. In contrast with the expressions presented in [12,18,19], Eq. (3) is valid for fractional values of $a$, which correspond to fractional number of subtracted photons caused by some experimental imperfections. Therefore, such parameterization is extremely effective for quantum state reconstruction.

In quantum mechanics the evolution of the density matrix under the action of losses can be obtained by solving the quantum-optical master equation [20]. In paper [12] the photon number distribution for damped MPSTS was obtained by interaction during the time $t$ with a thermal reservoir with a mean photon number $\mu_R$:

$$P_{\mu,a}(n,t) = \frac{[\mu_T(t)+1]^{a-1}[\mu e^{-\gamma t}+\mu_T(t)]^n}{[\mu e^{-\gamma t}+\mu_T(t)+1]^{a+n}} \cdot$$
$$\cdot {}_2F_1\left(1-a,-n;1;\frac{\mu e^{-\gamma t}}{[\mu_T(t)+1][\mu e^{-\gamma t}+\mu_T(t)]}\right), \quad (4)$$

where $\mu_T(t) = \mu_R(1-e^{-\gamma t})$, $\gamma$ – is a coupling constant and ${}_2F_1$ – is a Gaussian hypergeometric function. In the case of vacuum thermostat with $\mu_R = 0$ the coefficient $e^{-\gamma t}$ can be interpreted as an ordinary optical transmission coefficient. Then expression (4) can be substantially simplified and transformed again to the compound-Poisson distribution $P_{\mu(t),a}(n)$, where $\mu(t) = \mu e^{-\gamma t}$. In other words, optical losses don't change the coherence parameter $a$ of MPSTS, but decrease the mean photon number $\mu$ according to the transmission coefficient.

As an example, we present in Fig. 1 Wigner functions of multi-photon subtracted thermal states with parameters $a = 1, 2, 3$ and $\mu = 2, 4, 6$. Initial thermal states with $a = 1$ are Gaussian functions. For $a > 1$ the Wigner function becomes ring-shaped. It's non-Gaussianity rises with growth of $\mu$ and $a$. The $\hat{a}$ arrows correspond to the transformations due to the photon subtraction, which increase both parameters and lead to the increment of non-Gaussianity. Optical damping process is shown as the $\gamma t$ arrow. It conserves $a$, decrease $\mu$ and leads to the loss of non-Gaussianity.

However, we should underline, that optical damping doesn't transform MPSTS to the thermal state. Particularly, optical losses don't change the correlation function $g^{(2)} = 1+1/a$.

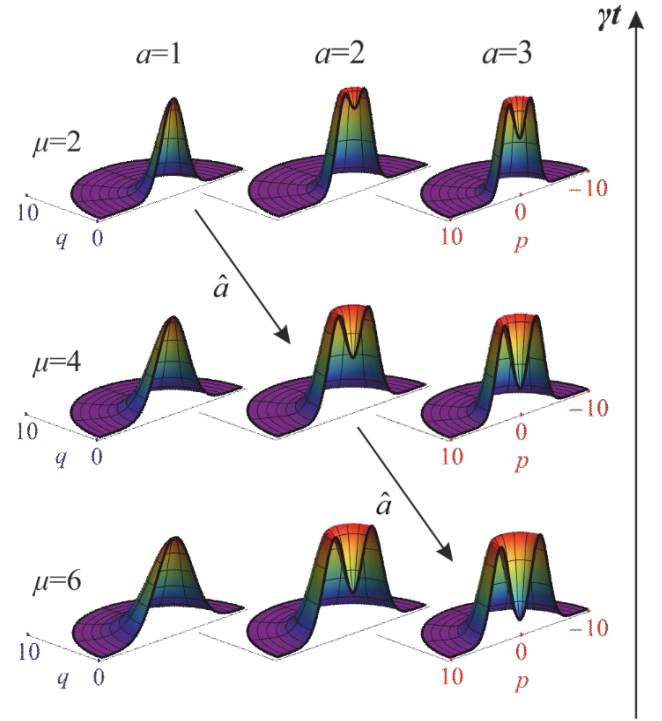

**Figure 1.** The Wigner functions of multi-photon subtracted thermal states with parameters $a = 1, 2, 3$ and $\mu = 2, 4, 6$. The $\hat{a}$ arrows correspond to the photon subtraction transformations and $\gamma t$ arrow shows the evolution during the optical damping.

## 3. Measures of non-Gaussianity

For the quantitative description of non-Gaussianity of MPSTSs, it is possible to introduce measures based on the distance between the density matrix $\hat{\rho}_{MPSTS}(\mu,a)$ and the density matrix of the closest Gaussian state, which is actually a thermal state with the same mean photon number $\hat{\rho}_{TS}(\mu)$.

One such measure is the Gilbert-Schmidt metric introduced to measure of non-Gaussianity in [21–23] and applied to the MPSTSs analysis in [12]:

$$\delta_{HS}(\hat{\rho}_{MPSTS}) \equiv \frac{\text{Tr}\left[(\hat{\rho}_{MPSTS} - \hat{\rho}_{TS})^2\right]}{2\text{Tr}\left[\hat{\rho}_{MPSTS}^2\right]} =$$
$$= \frac{1}{2}\left[1 + \frac{\sum_n \left(P_{\mu,1}^2(n) - 2P_{\mu,1}(n)P_{\mu,a}(n)\right)}{\sum_n P_{\mu,a}(n)}\right]. \quad (5)$$

The value of the measure $\delta_{HS}(\hat{\rho}_{MPSTS}) = 0$ if $\hat{\rho}_{MPSTS}$ is a Gaussian state. One should note that expression (5) can be represented in an analytical form [12].

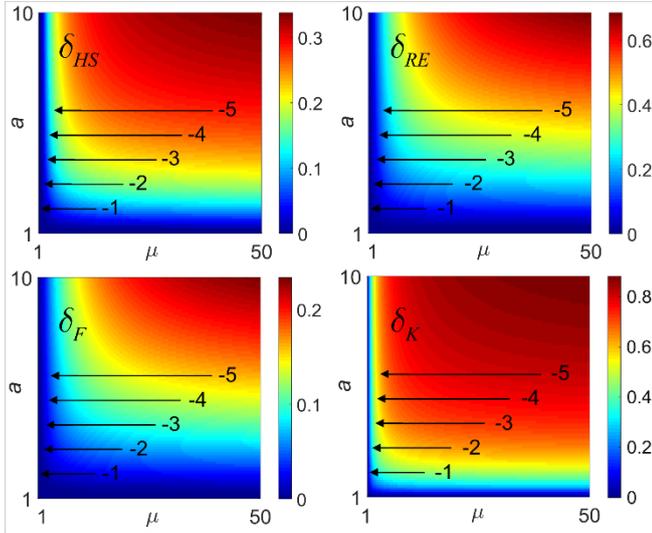

**Figure 2.** The density plots of the non-Gaussian measures for MPSTSs as a function of the parameters $a$ and $\mu$: $\delta_{HS}$ – the Hilbert-Schmidt measure (5), $\delta_{RE}$ – the relative entropy measure (6), $\delta_F$ – the Bures distance (7), $\delta_K$ – the measure based on the kurtosis of the quadrature distribution (12). Arrows correspond to the evolution trajectories for 1, 2, 3, 4 and 5-photon subtracted thermal states measured in the experiment.

The next measure that we are considering is the relative entropy metric [22,24]:

$$\delta_{RE}(\hat{\rho}_{MPSTS}) \equiv \text{Tr}\left[\hat{\rho}_{MPSTS}(\ln \hat{\rho}_{MPSTS} - \ln \hat{\rho}_{TS})\right] =$$
$$= (\mu+1)\ln(\mu+1) - \mu\ln\mu + \sum_n P_{\mu,a}(n)\ln P_{\mu,a}(n), \quad (6)$$

For the Gaussian thermal state $\delta_{RE}(\hat{\rho}_{TS}) = 0$.

The third measure is Bures distance [25,26], based on the Uhlmann's fidelity $F$ between the states $\hat{\rho}_{MPSTS}$ and $\hat{\rho}_{TS}$ [27,28]:

$$\delta_F(\hat{\rho}_{MPSTS}) \equiv 1 - \sqrt{F(\hat{\rho}_{MPSTS}, \hat{\rho}_{TS})} =$$
$$= 1 - \sum_n \sqrt{P_{\mu,1}(n)P_{\mu,a}(n)}, \quad (7)$$

where

$$F(\hat{\rho}_1, \hat{\rho}_2) = \left(\text{Tr}\left[\sqrt{\sqrt{\hat{\rho}_1}\hat{\rho}_2\sqrt{\hat{\rho}_1}}\right]\right)^2. \quad (8)$$

The disadvantage of all listed measures is the necessity to estimate the density matrix $\hat{\rho}$. Another way to define the non-Gaussianity is based on the shape of directly measurable by homodyne technique [29] quadrature distribution $P_{\mu,a}(q)$. It is based on the normalized fourth-order central moment – kurtosis [24]:

$$K \equiv \frac{m_4}{m_2^2}, \quad \beta_2 \equiv K - 3, \quad (9)$$

where $m_j$ is a $j$-order central moment and $\beta_2$ is called excess kurtosis. It characterizes the "flatness" of the distribution. For example, $\beta_2 = 0$ for Gaussian distribution, $\beta_2 = -1.2$ for uniform distribution, $\beta_2 = 3$ for Laplace distribution.

Within the model of the compound-Poisson photon number distribution, the quadrature distribution can be written as follows:

$$P_{\mu,a}(q) = \sum_n P_{\mu,a}(n)|\varphi_n(q)|^2, \quad (10)$$

where $\varphi_n(q)$ are the eigenfunctions of the harmonic oscillator. It turns out that one can obtain a simple analytic expression for the variance and excess kurtosis depending on the parameters $a$ and $\mu$ [15]:

$$m_2 = \mu + \frac{1}{2}, \quad \beta_2 = -6\left(\frac{\mu}{2\mu+1}\right)^2 \frac{a-1}{a}. \quad (11)$$

Then the measure of non-Gaussianity, normalized from 0 to 1, can be defined for MPSTS as

$$\delta_K(\hat{\rho}_{MPSTS}) \equiv \frac{2}{3}|\beta_2| = \left(\frac{2\mu}{2\mu+1}\right)^2 \frac{a-1}{a}. \quad (12)$$

The density plots for the above measures depending on the parameters $a$ and $\mu$ are presented in Fig. 2. It can be seen that distance-based measures $\delta_{HS}$, $\delta_{RE}$ and $\delta_F$ look similar: they smoothly increase with growth of $a$ and $\mu$. The kurtosis-based measure $\delta_K$, in contrast, rises sharply near the bounds of the plot and then continues to grow with much smaller gradient. So, we can conclude, that the distance-based measures work well on a large scale with $a, \mu \gg 1$, while the kurtosis-based measure is better to utilize on a small scale. The partial derivatives of all the measures with respect to $a$ and $\mu$ correspond to the sensitivity to the parameter estimation errors, which will be taken into account in the experimental part in Sec. 5.

## 4. Experimental verification

In order to observe the Gaussification of MPSTSs during the optical damping, the simple experimental setup was built. Its principal scheme is shown in Fig. 3. The HeNe cw laser radiation at the wavelength of 633 nm is asymmetrically distributed by the beamsplitter BS between two channels: 90% of the radiation serves as a homodyne (local oscillator field) and 10% is used for quantum state preparation.

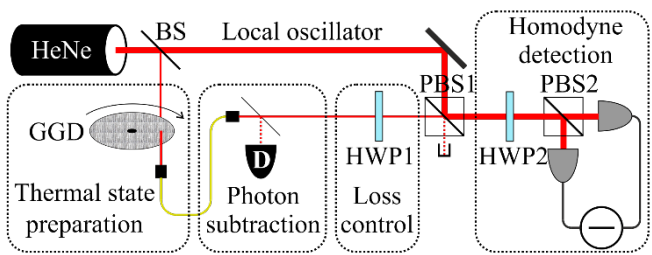

**Figure 3.** The scheme of the experimental setup for the preparation and measurement of damped MPSTSs. GGD – rotating ground glass disk, D – single-photon detector, BS – beam splitter, PBS – polarizing beam splitter, HWP – half wave plate.

The initial single-mode quasi-thermal light is prepared by passing laser beam throw a rotating ground glass disk GGD [30,31] followed by the single-mode fiber spatial filtering. The corresponding coherence time approximately equals to the time it takes for a grain of the disk to cross the beam and can be tuned by the disk velocity variation.

Conditional photon subtraction is realized by a beam splitter with reflectivity $r = 1\%$ [32] combined with an APD single photon detector Laser Components COUNT-100C-FC D with 100 Hz dark counts and a 50 ns dead time, placed in the reflection channel. It is necessary to note, that the coherence time of the thermal state was much greater than the APD dead time. This allows to detect serially several photocounts during the coherence time which corresponds to multiple photon subtraction – see [15] for details.

The optical loss is controlled by the rotation of the half-wave plate HWP1, followed by the polarized beam splitter PBS1. The homodyne radiation comes to the other input of PBS1.

Next both homodyne and MPSTS radiation leave PBS from the left output, second half-wave plate rotates polarization of both beams at 45° and they are symmetrically divided by the second polarization beam splitter PBS2 between two channels and come to the balanced homodyne detector Thorlabs PDB450A with a 100 kHz bandwidth and a 78% quantum efficiency.

Since the MPSTS Wigner function is axially symmetric, the homodyne phase wasn't varied during the quadrature distribution $P(q)$ measure.

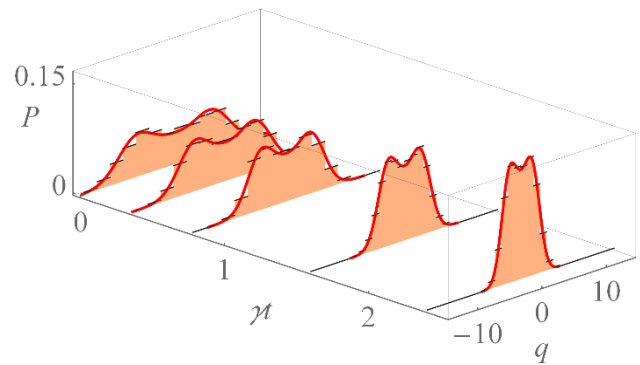

**Figure 4.** The measured quadrature histograms for 3-photon subtracted thermal state with initial mean photon number 34.6, at different levels of losses $\gamma t$ and correspondent fit $P(q)$.

The quantum state reconstruction was performed by the estimation of the parameters $a$ and $\mu$ from the measured quadrature histogram $P(q)$. It was done by the use of maximum likelihood technique and the model (10). The details can be also found in [15]. Such technique provides the extremal precision, because of small number of estimated parameters. Thus, model curve (10) always gave a good fit of the experimental histograms, that was checked by usual chi-2 test (see Fig. 4 for example).

For the estimation of confidence intervals for $a$ and $\mu$ we have used the Fisher information matrix [33,34].

The non-unitary quantum efficiency $\eta$ of the homodyne detector has been accounted as follows. It leads to the convolution of quadrature distribution $P(q)$ and normal probability distribution function with zero mean and variance $\sigma_c^2 = (1-\eta)/(2\eta)$ followed by the quadrature rescaling by the factor $\sqrt{\eta}$ [29]. This effect is equivalent to the transformation of the input quantum state under the additional losses $\eta$. So after the estimation of the parameters $a$ and $\mu$ from the raw quadrature data (without any rescaling and deconvolution) the mean photon number has been divided by $\eta$. But for the kurtosis measure one should take into account that the kurtosis excess is a normalized forth-order cumulant [35]. The cumulants additivity leads to the following equation between the ideal quadrature distribution kurtosis excess $\beta_2$ and the second and the fourth moments of the measured quadrature distribution $\tilde{m}_2$ and $\tilde{m}_4$:

$$\beta_2 \equiv \frac{m_4 - 3m_2^2}{m_2^2} = \frac{\tilde{m}_4 - 3\tilde{m}_2^2}{\left(\tilde{m}_2 - \eta\sigma_c^2\right)^2}. \quad (13)$$

## 5. Results and discussion

From 1- to 5-photon subtracted thermal states were prepared and measured under 5 levels of losses from zero to $\gamma t = 2.35$. The mean photon number for the initial thermal state without losses was 8.86.

The level of losses $\gamma t$ was determined from the mean photon number estimated from unconditional quadrature distribution.

For all the MPSTSs parameters $a$ and $\mu$ as well as their errors and covariations have been estimated and three measures of non-Gaussianity $\delta_{HS}$, $\delta_{RE}$ and $\delta_F$ have been calculated. The fourth measure $\delta_K$ has been calculated directly from the second and fourth moments of the measured quadrature histograms. The confidence intervals for $\delta_K$ values where determined by a numerical experiment.

The results are presented as points in Fig. 5. Theoretical values are presented as curves: bottom to top – one- to five- photon subtracted states at different loss levels $\gamma t$.

Here we should note, that due to some experimental imperfections the estimated coherence parameters didn't rich ideal values $a_M = M + 1$ for $M$-photon subtracted states. Measured values of $a$ equal 0.94, 1.61, 2.42, 3.27, 4.34, 5.46 for 0 to 5-subtracted states. So, we have used these values for theoretical curves in Fig. 5.

We can see, that experimental points lay close to the theoretical lines. To quantify the agreement between theoretical and experimental density matrices the fidelity $F(\hat{\rho}_{theory}, \hat{\rho}_{experiment})$ (8) has been calculated. For all the measured states it was higher than 99%.

However, one can note some disagreement between the theory and the experiment, which can be explained by the non-ideality of the photon subtraction process caused by dark-counts and instability of power and polarization of the laser radiation.

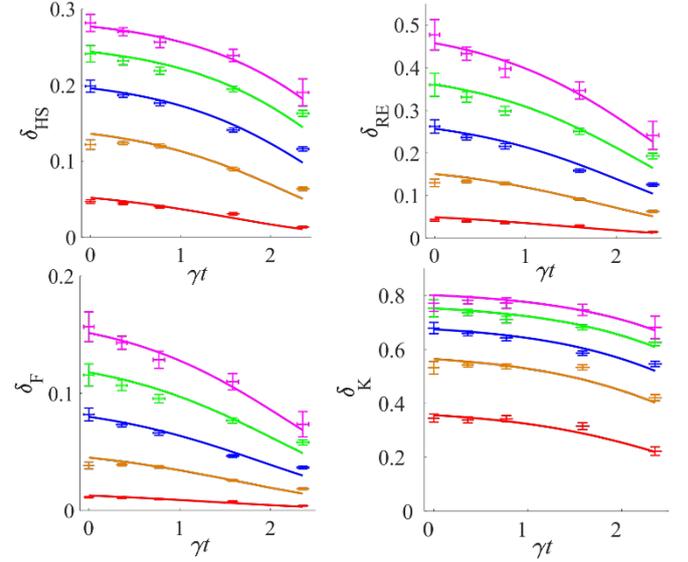

**Figure. 5.** Theoretical curves and experimental points for the measures of non-Gaussianity for different MPSTSs (bottom to top – one- to five-photon subtraction) at different loss levels $\gamma t$.

## 6. Conclusion

Thus, we have shown, how the several measures of non-Gaussianity depend on the parameters of MPSTS: mean photon number $\mu$ and coherence parameter $a$. Here we would like to highlight the kurtosis-based measure, which on the one hand have a simple analytical dependency on $\mu$ and $a$ (12), and on the other hand can be directly calculated from the measured quadrature histograms without quantum state reconstruction procedure (14). It has been shown theoretically and validated experimentally that MPSTSs lose their non-Gaussianity during the optical damping. The agreement between the theory and experiment reached 99%.

**Acknowledgments**

The work was supported by the Russian Science Foundation (RSF), project no.: 14-12-01338П.

**References**

[1] Eisert J, Scheel S and Plenio M B 2002 Distilling Gaussian States with Gaussian Operations is Impossible *Phys. Rev. Lett.* **89** 137903

[2] Fiurášek J 2002 Gaussian Transformations and Distillation of Entangled Gaussian States *Phys. Rev. Lett.* **89** 137904

[3] Giedke G and Ignacio Cirac J 2002 Characterization of Gaussian operations and distillation of Gaussian states *Phys. Rev. A* **66** 32316

[4] Browne D E, Eisert J, Scheel S and Plenio M B 2003 Driving non-Gaussian to Gaussian states with linear optics *Phys. Rev. A* **67** 62320

[5] Eisert J, Browne D E, Scheel S and Plenio M B 2004 Distillation of continuous-variable entanglement with optical means *Ann. Phys. (N. Y).* **311** 431–58

[6] Hage B, Franzen A, DiGuglielmo J, Marek P, Fiurášek J and Schnabel R 2007 On the distillation and purification of phase-diffused squeezed states *New J. Phys.* **9** 227–227

[7] Takahashi H, Neergaard-Nielsen J S, Takeuchi M, Takeoka M, Hayasaka K, Furusawa A and Sasaki M 2010 Entanglement distillation from Gaussian input states *Nat. Photonics* **4** 178–81

[8] Opatrný T, Kurizki G and Welsch D-G 2000 Improvement on teleportation of continuous variables by photon subtraction via conditional measurement *Phys. Rev. A* **61** 32302

[9] Cochrane P, Ralph T and Milburn G 2002 Teleportation improvement by conditional measurements on the two-mode squeezed vacuum *Phys. Rev. A* **65** 1–6

[10] Olivares S, Paris M G A and Bonifacio R 2003 Teleportation improvement by inconclusive photon subtraction *Phys. Rev. A* **67** 32314

[11] Cerf N J, Krüger O, Navez P, Werner R F and Wolf M M 2005 Non-Gaussian Cloning of Quantum Coherent States is Optimal *Phys. Rev. Lett.* **95** 70501

[12] Ghiu I, Marian P and Marian T A 2014 Loss of non-Gaussianity for damped photon-subtracted thermal states *Phys. Scr.* **T160** 14014

[13] Vidrighin M D, Dahlsten O, Barbieri M, Kim M S, Vedral V and Walmsley I A 2016 Photonic Maxwell's Demon *Phys. Rev. Lett.* **116** 1–7

[14] Hloušek J, Ježek M and Filip R 2017 Work and information from thermal states after subtraction of energy quanta *Sci. Rep.* **7** 13046

[15] Bogdanov Yu I, Katamadze K G, Avosopiants G V., Belinsky L V., Bogdanova N A, Kalinkin A A and Kulik S P 2017 Multiphoton subtracted thermal states: Description, preparation, and reconstruction *Phys. Rev. A* **96** 63803

[16] Bogdanov Yu I, Bogdanova N A, Katamadze K G, Avosopyants G V. and Lukichev V F 2016 Study of photon statistics using a compound Poisson distribution and quadrature measurements *Optoelectron. Instrum. Data Process.* **52** 475–85

[17] Bogdanov Yu I, Katamadze K G, Avosopyants G V, Belinsky L V, Bogdanova N A, Kulik S P and Lukichev V F 2016 Study of higher order correlation functions and photon statistics using multiphoton-subtracted states and quadrature measurements *SPIE Proceedings* vol 10224, ed V F Lukichev, E P Velikhov, A A Orlikovskiy, G Krasnikov and K V. Rudenko p 102242Q

[18] Allevi A, Andreoni A, Bondani M, Genoni M G and Olivares S 2010 Reliable source of conditional states from single-mode pulsed thermal fields by multiple-photon subtraction *Phys. Rev. A - At. Mol. Opt. Phys.* **82** 13816

[19] Zhai Y, Becerra F E, Glebov B L, Wen J, Lita A E, Calkins B, Gerrits T, Fan J, Nam S W and Migdall A 2013 Photon-number-resolved detection of photon-subtracted thermal light *Opt. Lett.* **38** 2171–3

[20] Breuer H-P and Petruccione F 2007 *The Theory of Open Quantum Systems* (Oxford University Press)

[21] Genoni M G, Paris M G A and Banaszek K 2007 Measure of the non-Gaussian character of a quantum state *Phys. Rev. A - At. Mol. Opt. Phys.* **76** 1–6

[22] Genoni M G, Paris M G A and Banaszek K 2008 Quantifying the non-Gaussian character of a quantum state by quantum relative entropy *Phys. Rev. A - At. Mol. Opt. Phys.* **78** 4–7

[23] Genoni M G and Paris M G A 2009 Non-Gaussianity and purity in finite dimensin *Int. J. Quantum Inf.* **7** 97–103

[24] Genoni M G and Paris M G A 2010 Quantifying non-Gaussianity for quantum information *Phys. Rev. A - At. Mol. Opt. Phys.* **82** 1–19

[25] Bures D 1969 An extension of Kakutani's theorem on infinite product measures to the tensor product of semifinite w∗-algebras *Trans. Am. Math. Soc.* **135** 199–199

[26] Ghiu I, Marian P and Marian T A 2013 Measures of non-Gaussianity for one-mode field states *Phys. Scr.* **87**

[27] Uhlmann A 1976 The "transition probability" in the state space of a ∗-algebra *Reports Math. Phys.* **9** 273–9

[28] Uhlmann A 1999 Fidelity and Concurrence of conjugated states *Phys. Rev. A* **62** 10

[29] Leonhardt U and Paul H 1995 Measuring the quantum state of light *Prog. Quantum Electron.*

**19** 89–130
[30] Martienssen W 1964 Coherence and Fluctuations in Light Beams *Am. J. Phys.* **32** 919
[31] Arecchi F T 1965 Measurement of the statistical distribution of gaussian and laser sources *Phys. Rev. Lett.* **15** 912–6
[32] Wenger J, Tualle-Brouri R and Grangier P 2004 Non-Gaussian Statistics from Individual Pulses of Squeezed Light *Phys. Rev. Lett.* **92** 153601
[33] Borovkov A A 1998 *Mathematical statistics* (New York: Gordon and Breach)
[34] Kendall M G and Stuart A 1979 *The Advanced Theory of Statistics. Inference and Relationship* (London: C. Griffin)
[35] Kendall M G and Stuart A 1977 *The Advanced Theory of Statistics. Distribution Theory* (London: C. Griffin)